\documentclass[12pt,thmsa,notitlepage,ukenglish,a4paper]{article}
\usepackage{amssymb}

\begin{document}

\author{\textbf{Jon Eakins and George Jaroszkiewicz} \\
School of Mathematical Sciences,University Park,\\
University of Nottingham, Nottingham NG7 2RD, UK}
\title{Factorization and Entanglement \\
in Quantum Systems}
\date{}
\maketitle

\begin{abstract}
We discuss the question of entanglement versus separability of pure quantum
states in direct product Hilbert spaces, and the relevance of this issue 
to physics.  Different types of separability may
be possible, depending on the particular factorization of the Hilbert space.
A given orthonormal basis set for a Hilbert space is defined to be of type
(p,q) if p elements of the basis are entangled and q are separable, relative
to a given bi-partite factorization of that space. We conjecture that not
all basis types exist for a given Hilbert space.
\end{abstract}

The phenomenon of entanglement is of central importance in the
interpretation of quantum mechanics. Historically, entanglement was the
focus of the famous Einstein-Podolsky-Rosen (\emph{EPR}) paper \cite{EPR},
which suggested that standard quantum mechanics is an incomplete theory of
physical reality. The central argument of the \emph{EPR} paper was that more
information about incompatible variables such as momentum and position could
in principle be deduced about an entangled two particle quantum state than
quantum mechanics permits, effectively giving information about each
particle separately, and therefore supporting a classical perspective.

The resolution of this ``paradox'' is the observation that information
extraction in quantum mechanics always comes at a cost: it is not possible
to actually extract information about incompatible variables from a given
state without destroying the state being looked at before the information
extraction process is completed, and this invalidates the argument used by
\emph{EPR }\cite{BOHR-35}\emph{. }

An apparently unrelated issue is the following. Throughout the history of
quantum mechanics, a constant topic of debate has been where the boundary
between the classical and quantum worlds should be. We believe that there is
now sufficient evidence to support the notion that there is no such
boundary, and that the classical world view is no more than an emergent,
i.e. effective, view of a universe which is entirely quantum mechanical in
origin \cite{JAROSZKIEWICZ-02A}. The evidence we cite is the near universal
validity of the quantized-field approach to elementary particles, numerous
experimentally observed violations of Bell inequalities and galactic
lensing. In the latter process, we can imagine observing (say) one photon
per day over several years to build up patterns analogous to the
interference bands seen in double slit experiments, the difference being
that the scale of the process is cosmological rather than local. In addition
to these more exotic applications, the validity of quantum principles is
supported by the overwhelming success of quantum mechanics in applied
physics, biology and chemistry, on both terrestrial and astrophysical scales.

With a recognition that the semi-classical observers of standard quantum
mechanics should ideally be regarded as quantum systems themselves, it has
become more fashionable to extend the quantum description to include them
with the systems under observation. This can be done whilst maintaining a
semi-classical perspective by writing a quantum state vector $\Psi $ for an
\emph{OS} (combined observer plus system under observation) as a direct
product $\Psi \equiv \theta \otimes \phi $, where $\theta $ represents a
state of the observer \emph{O }and $\phi $ represents a state of the system
\emph{S }under observation. Such a state will be called \emph{separable.} In
general, we shall use the word \emph{separable} when we talk about states
constructed from direct products of vectors, and \emph{factorizable} when we
refer to Hilbert spaces constructed from direct (tensor) products of
(factor) Hilbert spaces. When applied to Hilbert spaces, the term \emph{%
separable} traditionally refers to the possibility of finding a countable
basis for it, regardless of any issue of factorizability.

When the dimensions of realistic Hilbert spaces which might model the
universe are considered, then the \emph{a priori} probability that a state
chosen at random in such a space be separable is zero, as will be seen from
our discussion of concurrency below. The observed separability of the
universe into vast numbers of identifiably distinct subsystems is on the
face of it surprising from this point of view. However, this does not take
into account the crucial role of dynamics, which imposes very specific
constraints on which states are physically accessible in the course of time.
For example, suppose all the possible outcomes of some quantum process are
separable states. Then there will be zero probability of getting an
entangled state outcome in that process.

In this article, classicity (or classicality) is regarded as synonymous with
the possibility of making distinctions between different objects, such as
different spatial positions, or physical subsystems. In quantum mechanics,
entanglement may be regarded as a breakdown of such a possibility. When
physicists discuss isolated systems within a wider universe, they invariably
model the totality by separable states, with some of the factors
representing states of the isolated systems and other factors representing
the rest of the universe. The conventional procedure is then to ignore these
other factors (the environment), and discuss only those factors representing
the isolated systems. Certainly, it seems impossible to discuss experiments
in physics without assuming that the states of interest are factored out
from the rest of the environment. The development of decoherence theory has
not altered this in the least. Separability is therefore as fundamental to
quantum physics as entanglement.

This leads to the following question: given a finite dimensional Hilbert
space $\mathcal{H}$ of dimension $d\equiv \dim \mathcal{H}$, when is it
possible to think of a state $\Psi $ in $\mathcal{H}$ as a separable state?
By this we mean we would like to know the circumstances which guarantee that
$\Psi $ is a tensor product of the form $\Psi =\psi \otimes \phi ,$ where $%
\psi $ is some vector in some factor space $\mathcal{H}_{1}$ of $\mathcal{H}$
and $\phi $ is another vector in another factor $\mathcal{H}_{2}$ of $%
\mathcal{H}$.

Let $\mathcal{H}$ be a finite dimensional Hilbert space of dimension $\dim
\mathcal{H}$. If $\mathcal{H}$ can be expressed in the bi-partite form
\begin{equation}
\mathcal{H}=\mathcal{H}_1^{(d_1)}\otimes \mathcal{H}_2^{(d_2)},  \label{889}
\end{equation}
where $\mathcal{H}_i^{(d_i)},i=1,2$ is a Hilbert space with dimension $d_i,$
then we shall say that $\mathcal{H}$ is \emph{factorizable. }Clearly $\dim
\mathcal{H}$ must itself be factorizable and given by the rule $\dim
\mathcal{H}=d_1d_2.$

This elementary result may have important cosmological implications.
According to a number of authors [3-6] the universe is described by a time
dependent pure quantum state $\Psi $, an element in a Hilbert space $%
\mathcal{H}_U$ of enormous but finite dimension. We note that the notion
that the universe is a quantum system has been criticised principally on the
grounds that there is no evidence that all physical systems must possess
quantum states \cite{FINK-00}, and also because it appears inconsistent to
discuss probabilities when there is only one universe. These arguments can
be met with three counter arguments: first, the absence of any boundary
between the quantum and classical worlds and the empirical validity of
quantum mechanics actually strongly supports the notion that \emph{all}
systems must run on quantum principles, and so by extension does the
universe; second, a pure state formalism eliminates the need for a density
matrix approach to quantum cosmology; third, quantum probabilities make
sense if they are interpreted correctly in terms of \emph{predictions} about
the possible future state of the universe made by physicists who are
themselves part of the quantum universe. This is not inconsistent with the
notion that the universe is in a definite state in the present.

Given this quantum perspective about the universe, the apparently
overwhelmingly classical appearance of the universe, with a classical
looking spatial structure which permits the separation of vast numbers of
subsystems of the universe spatially, is interpreted by us as evidence that
the current state of the universe $\Psi $ has separated into a vast number
of factors. If this is true then $\mathcal{H}_U$ must be factorizable into a
vast number of factor spaces, and therefore $\dim \mathcal{H}_U$ itself must
be highly factorizable. In particular, the Hilbert space of the universe
cannot have prime dimension according to this scenario.

Hilbert spaces with a high degree of factorizability are readily
constructed. Recent approaches to fundamental physics inspired by spin
networks and quantum computation [5,8] considers $\mathcal{H}_U$ to be the
direct product of a (usually vast) number $N$ of qubit Hilbert spaces, viz
\begin{equation}
\mathcal{H}_U=\mathcal{H}_1^{(2)}\otimes \mathcal{H}_2^{(2)}\otimes \ldots
\otimes \mathcal{H}_N^{(2)},  \label{567}
\end{equation}
and then $\dim {\mathcal{H}}_{U} =2^N.$ None of the individual qubit factor
spaces $\mathcal{H}_i^{(2)}$ are factorizable, so that $\left( \ref{567}%
\right) $ represents a complete, or maximal, factorization of $\mathcal{H}_U$%
. We shall call such a qubit factorization a \emph{primordial factorization.
}In the most general case, a primordial factorization will be of the form
\begin{equation}
\mathcal{H=H}_1^{(p_1)}\otimes \mathcal{H}_2^{(p_2)}\otimes \ldots \otimes
\mathcal{H}_N^{(p_N)},
\end{equation}
where the $p_i$ are prime numbers and $\dim \mathcal{H}=p_1p_2\ldots p_N$.
If the \emph{factorizability }$\zeta $ of $\mathcal{H}$ is defined as the
ratio $N/\dim \mathcal{H}$ then qubits provide the maximum factorizability
for a given $N$, i.e., $\zeta =N/2^N.\;$Qubits are favoured by various
authors because they represent the most elementary attributes of logic, that
is, ``yes'' and ``no'' (or equivalently, ``true'' and ``false'') can be
identified with the two elements of a qubit ``spin-up'', ``spin-down'' basis.

Given an $N-$ qubit system with $N>2$ then it is possible to consider \emph{%
partial factorizations} (or \emph{splits}) of $\mathcal{H}$ of the form
\begin{equation}
\mathcal{H=H}^{(2^{n})}\otimes \mathcal{H}^{(2^{N-n})},  \label{346}
\end{equation}
where
\begin{eqnarray}
\mathcal{H}^{(2^{n})} &\equiv &\mathcal{H}_{1}^{(2)}\otimes \mathcal{H}%
_{2}^{(2)}\otimes \ldots \otimes \mathcal{H}_{n}^{(2)},  \nonumber \\
\mathcal{H}^{(2^{N-n})} &\equiv &\mathcal{H}_{n+1}^{(2)}\otimes \mathcal{H}%
_{n+2}^{(2)}\otimes \ldots \otimes \mathcal{H}_{N}^{(2)},
\end{eqnarray}
and variants of this theme. If a given partial factorization has two
factors, such as in (\ref{346}) we shall call this a \emph{bi-partite
factorization.}

\ $\;$

We now discuss a necessary and sufficient condition for the separability of
a state relative to a given bi-partite factorization.

Let $\mathcal{B}_a^{(d_a)}\equiv \left\{ |i\rangle _a:i=1,\ldots
,d_a\right\} $ be an orthonormal basis for factor space $\mathcal{H}%
_a^{(d_a)}.$\ Orthonormality is not necessary for our theorem below to hold
but is useful in subsequent discussions. Given that $\mathcal{H}$ is
factorizable in the form (\ref{889}) then an orthonormal basis for $\mathcal{%
H}$ is $\mathcal{B}\equiv \{|i\rangle _1\otimes |j\rangle _2:i=1,2,\ldots
,d_1,\;j=1,2,\ldots d_2\}.\;$ Such a basis will be called a \emph{%
factorizable basis. }Any state $|\Psi \rangle $ in $\mathcal{H}$ can then be
written in the form
\begin{equation}
|\Psi \rangle =\sum_{i=1}^{d_1}\sum_{j=1}^{d_2}C_{ij}|i\rangle _1\otimes
|j\rangle _2,  \label{aaa}
\end{equation}
where the coefficients $C_{ij}$ are complex and form the components of a
complex $d_1\times d_2\;$matrix called the \emph{coefficient matrix.} It is
relatively easy to prove the following theorem:

\subparagraph{Theorem:}

The state $|\Psi \rangle $ is separable \emph{relative to the factorizable
basis }$\mathcal{B}$ if and only if the coefficient matrix satisfies the
\emph{micro-singularity }condition\
\begin{equation}
C_{ij}C_{ab}=C_{ib}C_{aj}  \label{991}
\end{equation}
for all possible values of the indices. \emph{\ }A proof involving the
concept of \emph{concurrency} is given in \cite{ALBEVERIO-01}. For example,
a state $|\Psi \rangle $ in a two-qubit system\ of the form
\begin{eqnarray}
|\Psi \rangle &=&{}\alpha |1\rangle _1\otimes |1\rangle _2+\beta |1\rangle
_1\otimes |2\rangle _2+  \nonumber \\
&&\gamma |2\rangle _1\otimes |1\rangle _2+\delta |2\rangle _1\otimes
|2\rangle _2
\end{eqnarray}
is separable if and only if {}$\alpha \delta =\beta \gamma $, which can be
readily verified.

\smallskip There are two points to make here. First, given a factorizable
basis, a coefficient matrix chosen at random will almost certainly not be
micro-singular, simply because for large dimensions, there will be a vast
number of micro-singularity conditions $\left( \ref{991}\right) $ to
satisfy. The number $N_{C}$ of such conditions will in general be given by $%
N_{C}=\frac{_{1}}{^{4}}d_{1}\left( d_{1}-1\right) d_{2}\left( d_{2}-1\right)
\sim \frac{_{1}}{^{4}}{dim}^{2}\mathcal{H}$ for large $dim\mathcal{H}$.\
This is why the existence of separability in a universe which is running on
quantum principles should come as a surprise. Rather than envisage
entanglement as an extraordinary phenomenon, we should perhaps ask why the
degree of separability in the current epoch of the universe is so relatively
large. We envisage that, given a fully quantum universe which was jumping
from one quantum state to another, most of these states should be entangled,
unless there is some very special reason for separability. A related issue
is the idea, consistent with recent developments in quantum gravity, that
space itself is an emergent attribute of a completely quantum universe
[3,10]. It is hard to understand how this attribute could emerge unless
successive states of the universe in the current epoch were highly separable
and remained so under the influence of extraordinary dynamical laws. Without
separability, there can be no notion of classicity, and without any form of
classicity, the concept of space itself cannot be formulated. Position in
space is, after all, synonymous with the classical statement that \emph{this}
object is \emph{here} and \emph{not} \emph{there}. As the \emph{EPR }%
discussion shows, such a statement is not always possible for entangled
states. Moreover, from this viewpoint, the expansion of the universe may be
taken as some indicator that, far from being of very low probability,
separability is actually increasing, suggesting that the current dynamics of
the universe is somehow organizing a greater degree of \emph{classicity} (or
separability) with time.

To illuminate the scale of the problem of explaining the current
separability of the universe, a simple estimate of the lowest realistic
dimension $d_{U}$ of the Hilbert space $\mathcal{H}_{U}$ of the universe
gives $d_{U}\gtrsim 2^{10^{180}},\;$which is based on the supposition that
each Planck volume in the visible universe contains one elementary qubit.
More realistic estimates would certainly increase this estimate
dramatically. The set of separable states in $\mathcal{H}_{U}$ is a set of
measure zero, with the number of concurrency conditions being proportional
to $d_{U}^{2}$ for large dimensions$.$ We should ask, therefore, why does
the universe have so much apparent separability, that is, why can physicists
investigate isolated systems at all?

The second point is that separability depends on the choice (if any) of the
factorization of the total Hilbert space $\mathcal{H}$. Consider the Hilbert
space $\mathcal{H}\equiv \mathcal{H}_1^{(2)}\otimes \mathcal{H}%
_2^{(2)}\otimes \mathcal{H}_3^{(2)}$ where each $\mathcal{H}_i^{(2)}$
represents a qubit space. Now rearrange $\mathcal{H}$ in the form bi-partite
form
\begin{equation}
\mathcal{H=H}_A^{(4)}\otimes \mathcal{H}_3^{(2)},
\end{equation}
where $\mathcal{H}_A^{(4)}\equiv \mathcal{H}_1^{(2)}\otimes \mathcal{H}%
_2^{(2)}$,\ with factorizable basis
\begin{equation}
\mathcal{B}_1\equiv \{|ij\rangle _A\otimes |k\rangle _3:\;1\leqslant
i,j,k\leqslant 2\}.
\end{equation}
Then a separable state $|\Psi \rangle $ relative to this factorization and
this basis will be of the form
\begin{equation}
|\Psi \rangle =\left( a|11\rangle _A+b|12\rangle _A+c|21\rangle
_A+d|22\rangle _A\right) \otimes ({}\alpha |1\rangle _3+\beta |2\rangle _3),
\end{equation}
where the coefficients $a,b,\ldots ,\beta $ are complex, giving the
coefficient matrix
\begin{equation}
\begin{array}{ccc}
\otimes & \mathbf{1}_3 & \mathbf{2}_3 \\
\mathbf{11}_A & a{}\alpha & a\beta \\
\mathbf{12}_A & b{}\alpha & b\beta \\
\mathbf{21}_A & c{}\alpha & c\beta \\
\mathbf{22}_A & d{}\alpha & d\beta
\end{array}
,
\end{equation}
which clearly satisfies micro-singularity. In this matrix, the top row and
left-most column label basis vectors for the different factor spaces, and
the other terms represent the coefficients of their tensor products, i.e.,
the actual elements of the coefficient matrix.

Now we may also write the Hilbert space in the alternative bi-partite form
\begin{equation}
\mathcal{H}=\mathcal{H}_{1}^{(2)}\otimes \mathcal{H}_{B}^{(4)},
\end{equation}
where\ $\mathcal{H}_{B}^{(4)}\equiv \mathcal{H}_{2}^{(2)}\otimes \mathcal{H}%
_{3}^{(2)}.\;$We note for example
\begin{equation}
|12\rangle _{A}\otimes |1\rangle _{3}=|1\rangle _{1}\otimes |21\rangle _{B}
\end{equation}
and so on. Then the above state is given by
\begin{equation}
|\Psi \rangle =a{}\alpha |1\rangle _{1}\otimes |11\rangle _{B}+a\beta
|1\rangle _{1}\otimes |12\rangle _{B}+...
\end{equation}
giving the alternative coefficient matrix
\begin{equation}
\begin{array}{ccccc}
\otimes & \mathbf{11}_{B} & \mathbf{12}_{B} & \mathbf{21}_{B} & \mathbf{22}%
_{B} \\
\mathbf{1}_{1} & a{}\alpha & a\beta & b{}\alpha & b\beta \\
\mathbf{2}_{1} & c{}\alpha & c\beta & d{}\alpha & d\beta
\end{array}
,
\end{equation}
which clearly does not satisfy micro-singularity. Therefore, a state
separable relative to one factorization of $\mathcal{H}$ is not necessarily
separable relative to another factorization of $\mathcal{H}$. We see here a
basic difference between the mathematical and physical
descriptions of states. Mathematicians tell us that separable
and entangled states can be transformed into each other, whereas
physicists regards their differences as physically significant in
the right context.

Apart from having consequences for the question of separability on emergent
scales, this may have another cosmological implication. Given that the
separability of the state of the universe is meaningful only relative to a
special factorization of $\mathcal{H}$, this suggests that there is a \emph{%
preferred} basis for $\mathcal{H}$ before each jump.\ If the universe can
jump only into an eigenstate of some specific complete set of observables
\cite{JAROSZKIEWICZ-02A}, then that preferred basis will be the set of
possible eigenstates (i.e. the possible outcomes) of that complete set of
observables. This complete set may change with each jump, but nevertheless
this picture must still hold. There must be something extraordinarily
special about the selection of this set of observables. For instance, in the
current epoch, the possible outcomes appear highly separable. In a fully
quantized universe running as a quantum automaton, this choice cannot be
made by any external agency. Given that the number of independent Hermitian
operators is of the order $d_{\mathcal{H}}^{2}$ \cite{PERES:93}, then there
must be some as yet unknown and very specific laws which determine the
operators responsible for the separability of the universe in the current
epoch.

$\;$

An important feature of quantum theory is that individual elements of a
factorizable space are not in general separable relative to a primordial
factorization. If $\psi $ is an element of $\mathcal{H}_{1}^{(d_{1})}$ and $%
\phi $ is an element of $\mathcal{H}_{2}^{(d_{2})}$, then $\Psi \equiv \psi
\otimes \phi $ is a separable element of $\mathcal{H}\equiv \mathcal{H}%
_{1}^{(d_{1})}\otimes \mathcal{H}_{2}^{(d_{2})}$.\ We shall say that $\Psi $
\emph{is separable relative to the} $(\mathcal{H}_{1}^{(d_{1})},\mathcal{H}%
_{2}^{(d_{2})})$ \emph{factorization of} $\mathcal{H}$. It is a particularly
important fact that the separability of $\Psi $ relative to $(\mathcal{H}%
_{1}^{(d_{1})},\mathcal{H}_{2}^{(d_{2})})$ does not depend on the choice of
basis for $\mathcal{H}_{1}^{(d_{1})}$ or for $\mathcal{H}_{2}^{(d_{2})}$,
that is, this separation is invariant to (local) unitary transformations of
basis for $\mathcal{H}_{1}^{(d_{1})}$ and $\mathcal{H}_{2}^{(d_{2})}$%
separately. If an element $\Theta $ of $\mathcal{H}$ is not \emph{separable
relative to a} $(\mathcal{H}_{1}^{(d_{1})},\mathcal{H}_{2}^{(d_{2})})$ \emph{%
factorization of} $\mathcal{H}$ then we shall say that $\Theta $ is \emph{%
entangled} \emph{relative to the} $(\mathcal{H}_{1}^{(d_{1})},\mathcal{H}%
_{2}^{(d_{2})})$ \emph{factorization of} $\mathcal{H}.$ Such a statement
would also be invariant to local transformations of basis for $\mathcal{H}%
_{1}^{(d_{1})}$ and $\mathcal{H}_{2}^{(d_{2})}$separately.

Now consider an arbitrary orthonormal basis $\mathcal{B}$ for $\mathcal{H}$.
This will have $d\equiv d_{1}d_{2}$ elements $\beta _{i},i=1,2,\ldots ,d.$
The question we ask now is, how many of these are separable relative to the
factorization $(\mathcal{H}_{1}^{(d_{1})},\mathcal{H}_{2}^{(d_{2})})$ of $%
\mathcal{H}$ and how many are entangled? If $q$ of the $\beta _{i}$ are
separable and $p=d-q$ are entangled, that is, not separable relative to $(%
\mathcal{H}_{1}^{(d_{1})},\mathcal{H}_{2}^{(d_{2})}),$ then we shall say
that $\mathcal{B}$ \emph{is of type }$(p,q).$ If $q=d$ then $\mathcal{B}$ is
\emph{a completely separable basis relative to the factorization }$(\mathcal{%
H}_{1}^{(d_{1})},\mathcal{H}_{2}^{(d_{2})})$ whereas if $p=d$ then $\mathcal{%
B}$ is \emph{a completely entangled basis} \emph{relative to this
factorization. }Otherwise, $\mathcal{B}$ is a \emph{partially separable (or
partially entangled) basis relative to the factorization} $(\mathcal{H}%
_{1}^{(d_{1})},\mathcal{H}_{2}^{(d_{2})}).$

In such a bipartite factorization, each of the factor spaces $\mathcal{H}%
_1^{(d_1)},\mathcal{H}_2^{(d_2)}$ could in principle be factorizable into
two or more factors, and this would then lead to a natural extension of this
sort of classification of bases, which is left to the reader to explore.

We now discuss an example which suggests that not every type of partially
factorizable basis exists. This example is a two qubit system, so that $d=4$%
, and $\mathcal{H}$ has the primordial factorization
\begin{equation}
\mathcal{H=H}_1^{(2)}\otimes \mathcal{H}_2^{(2)},
\end{equation}
where the $\mathcal{H}_i^{(2)}$, $i=1,2$ are individual two-dimensional
qubit Hilbert spaces. Without loss of generality we shall work with a
specific choice of basis for each factor Hilbert space. For qubit $%
i,\{|0\rangle _i,|1\rangle _i:i=1,2\}$ is an orthonormal basis for $\mathcal{%
H}_i^{(2)}.$ Then we define
\begin{equation}
|ij\rangle \equiv |i\rangle _1\otimes |j\rangle _2,\;\;\;\;\;\;0\leqslant
i,j\leqslant 1.
\end{equation}

We shall give examples of type $\left( 0,4\right) $, $\left( 2,2\right)
,\;\left( 3,1\right) $ and $\left( 4,0\right) $ bases for $\mathcal{H}$, and
then a proof that type $\left( 1,3\right) $ does not exist.

\subparagraph{Type $(0,4)$}

With the above notation, a completely factorizable basis, that is, a type $%
\left( 0,4\right) $ basis $\mathcal{B}_{0,4}$ is given by
\begin{equation}
\mathcal{B}_{0,4}=\left\{ |00\rangle ,|01\rangle ,|10\rangle ,|11\rangle
\right\} .
\end{equation}

\subparagraph{Type $(2,2)$}

With the same notation as above, a type $\left( 2,2\right) $ basis for $%
\mathcal{H}$ is given by
\begin{equation}
\mathcal{B}_{2,2}=\left\{ |00\rangle ,|11\rangle ,\frac{_1}{^{\sqrt{2}}}%
\left( |01\rangle +|10\rangle \right) ,\frac{_1}{^{\sqrt{2}}}\left(
|01\rangle -|10\rangle \right) \right\} .
\end{equation}

\subparagraph{Type $\left( 3,1\right) $}

$\;$Relative to the basis $\mathcal{B}_{0,4}$ given above, a type $\left(
3,1\right) $ orthonormal basis for $\mathcal{H}$ is given by
\begin{equation}
\mathcal{B}_{3,1}=\left\{ |00\rangle ,\frac{_1}{^{\sqrt{2}}}|11\rangle +%
\frac{_1}{^2}|01\rangle +\frac{_1}{^2}|10\rangle ,\frac{_1}{^{\sqrt{2}}}%
|11\rangle -\frac{_1}{^2}|01\rangle -\frac{_1}{^2}|10\rangle ,\frac{_1}{^{%
\sqrt{2}}}\{|01\rangle -|10\rangle \}\right\}
\end{equation}

\subparagraph{Type $\left( 4,0\right) $}

$\;$Relative to the basis $\mathcal{B}_{0,4}$ given above, a type $\left(
4,0\right) $ orthonormal basis for $\mathcal{H}$ is given by
\begin{equation}
\mathcal{B}_{3,1}=\left\{ \frac{_1}{^{\sqrt{2}}}\{|00\rangle +|11\rangle \},%
\frac{_1}{^{\sqrt{2}}}\{|00\rangle -|11\rangle \},\frac{_1}{^{\sqrt{2}}}%
\{|01\rangle +|10\rangle \},\frac{_1}{^{\sqrt{2}}}\{|01\rangle -|10\rangle
\}\right\}
\end{equation}

The existence of type $(3,1)$ partially factorizable bases makes it
surprising that no type $\left( 1,3\right) $ basis can exist. Although
intuitively obvious, a proof is surprisingly long:

\subparagraph{Theorem:}

No type $\left( 1,3\right) $ basis of a two-qubit Hilbert space exists
relative to the primordial factorization.

\subparagraph{Proof:}

Let $\eta _{1,}\eta _2,$ and $\eta _3$ be three mutually orthogonal vectors
which are separable relative to the primordial factorization $\mathcal{%
H\equiv H}_1^{(2)}\otimes \mathcal{H}_2^{(2)}$ of a two qubit Hilbert space $%
\mathcal{H}$. By definition these vectors are of the form
\begin{equation}
\eta _i=\psi _i\otimes \phi _i,\;\;\;i=1,2,3,
\end{equation}
where $\psi _i\in \mathcal{H}_1^{(2)}$ and $\phi _i\in \mathcal{H}_2^{(2)}.$
None of the factor vectors $\psi _i,\phi _i$ can be zero, since we require
\begin{equation}
(\eta _i,\eta _i)=(\psi _i,\psi _i)_1(\phi _i,\phi _i)_2>0,\;\;\;i=1,2,3,
\end{equation}
where subscripts on inner products refer to the corresponding factor space.

Mutual orthogonality give the three conditions
\begin{eqnarray}
(\psi _1,\psi _2)_1(\phi _1,\phi _2)_2 &=&0  \nonumber \\
(\psi _1,\psi _3)_1(\phi _1,\phi _3)_2 &=&0  \label{234} \\
(\psi _2,\psi _3)_1(\phi _2,\phi _3)_2 &=&0.  \nonumber
\end{eqnarray}

Now define $A_{ij}\equiv (\psi _i,\psi _j)_1$,$\;B_{ij}\equiv (\phi _i,\phi
_j)_2$ for $1\leqslant i<j\leqslant 3.$ First, we show that not all three of
the $A_{ij}$ can be zero. Suppose this were true. Then
\begin{equation}
A_{12}=0 \Rightarrow (\psi _1,\psi _2)_1=0.\;
\end{equation}
Since none of the $\psi _i$ can be zero and since $\mathcal{H}_1^{(2)}$ is
two dimensional, we deduce that $\psi _1$ and $\psi _2$ form an orthogonal
basis for $\mathcal{H}_1^{(2)}$. Hence we may write
\begin{equation}
\psi _3=a\psi _1+b\psi _2,\;\;\;|a|^2+|b|^2>0.  \label{123}
\end{equation}
Then
\begin{eqnarray}
A_{13} &=&0\Rightarrow a=0,  \nonumber \\
A_{23} &=&0\Rightarrow b=0,
\end{eqnarray}
which contradicts $\left( \ref{123}\right) .$ Likewise, not all the $B_{ij}$
can be zero.

Without loss of generality, the only way to satisfy the mutual orthogonality
conditions (\ref{234}) is to have $A_{12}=A_{13}=B_{23}=0$ and $A_{23}\neq 0$%
.\ (By symmetry, any other choice of the $A_{ij}$, $B_{jk}$ would be just as
good).

With these conditions assumed, we use the condition $A_{12}=0$ as before to
deduce condition (\ref{123}). The condition $A_{13}=0$ gives $a=0$ and so we
deduce $b\neq 0.$

Similarly, $B_{23}=0$ implies $\phi _2$ and $\phi _3$ form an orthogonal
basis for $\mathcal{H}_2^{(2)}$, and therefore we may write
\begin{equation}
\phi _1=c\phi _2+d\phi _3,\;\;\;\;\;|c|^2+|d|^2>0.
\end{equation}
Hence we may write
\begin{eqnarray}
\eta _1 &=&\psi _1\otimes (c\phi _2+d\phi _3),  \nonumber \\
\eta _2 &=&\psi _2\otimes \phi _2 \\
\eta _3 &=&b\psi _2\otimes \phi _3.  \nonumber
\end{eqnarray}

With these results we see for example that a completely factorizable
orthogonal basis, i.e., a type (0,4) basis, for $\mathcal{H}$ is given by
\begin{equation}
\mathcal{B}_{0,4}=\left\{ \psi _1\otimes \phi _2,\psi _1\otimes \phi _3,\psi
_2\otimes \phi _2,\psi _2\otimes \phi _3\right\} .
\end{equation}
Now consider a non-zero vector $\eta _4.$ This may be written in the form
\begin{equation}
\eta _4={}\alpha \psi _1\otimes \phi _2+\beta \psi _1\otimes \phi _3+\gamma
\psi _2\otimes \phi _2+\delta \psi _2\otimes \phi _3
\end{equation}
with
\begin{equation}
|{}\alpha |^2+|\beta |^2+|\gamma |^2+|\delta |^2>0.
\end{equation}
If the $\eta _i$ form an orthogonal, type $(1,3)$ basis then we must have
\begin{equation}
(\eta _1,\eta _4)=(\eta _2,\eta _4)=(\eta _3,\eta _4)=0,
\end{equation}
plus the micro-singularity condition discussed above that $\eta _4$ is
entangled relative to the primordial basis, i.e.
\begin{equation}
C_4\equiv {}\alpha \delta -\beta \gamma \neq 0.  \label{456}
\end{equation}
The orthogonality conditions give
\begin{eqnarray}
(\eta _2,\eta _4) &=&0\;\;\;\Rightarrow \gamma =0,  \nonumber \\
(\eta _3,\eta _4) &=&0\;\;\;\Rightarrow \delta =0.
\end{eqnarray}
This gives
\begin{equation}
C_4=0,
\end{equation}
however, which is inconsistent with (\ref{456}), and hence the theorem is
proved.

$\;$

The above discussion of bases has some implications concerning operators. If
$\hat{A}$ is any Hermitian operator on $\mathcal{H}$ with non-degenerate
eigenvalues then there is a unique orthonormal basis $\mathcal{B}_{A}$ for $%
\mathcal{H}$ formed from the normalized eigenvectors $\psi _{{}{}\alpha }$
of $\hat{A}$ \cite{PERES:93}. Suppose that $\mathcal{H}$ is factorizable
with bi-partite factorization
\begin{equation}
\mathcal{H}=\mathcal{H}_{1}^{(d_{1})}\otimes \mathcal{H}_{2}^{(d_{2})},\;\;%
\;d_{1},d_{2}>1.  \label{178}
\end{equation}
Then $\mathcal{B}_{A}$ will be of type $(r,s)$ relative to this
factorization, where $r+s=d_{1}d_{2}$, for some non-negative integers $r$
and $s$. For any orthonormal basis for $\mathcal{H}$, such a classification
is unique, relative to the given bi-partite factorization, and therefore\ we
conclude that any non-degenerate Hermitian operator on $\mathcal{H}$ can be
assigned a unique classification $(r,s),$ relative to a given bi-partite
factorization of $\mathcal{H}$.

A conclusion from our previous result is that there are no type $\left(
1,3\right) $ Hermitian operators acting on elements of a two qubit system.

If a Hermitian operator $\hat{A}$ has two or more degenerate eigenvalues
then there is no uniqueness in the construction of an orthonormal basis from
its eigenvectors. In such a case, more Hermitian operators $\hat{B}$, $%
\hat{C},\ldots $ commuting with $\hat{A}$ and with each other need to be
found in order to form a complete commuting set \cite{PERES:93}\ $\mathcal{S}%
\equiv \left\{ \hat{A},\hat{B},\hat{C},\ldots \right\} .\;$A complete
commuting set then gives a unique orthonormal basis $\mathcal{B}_{S}$ for $%
\mathcal{H}$, which will be of unique type $\left( r,s\right) $ relative to
the factorization (\ref{178}). Hence we can classify uniquely any complete
commuting set as being of type $\left( r,s\right) $ relative to a given
factorization (\ref{178}) of the Hilbert space. This will be an important
classification in physical situations involving physically identifiable
factor spaces, such as qubit registers in quantum computers.

\paragraph{\protect\Large Concluding remarks:}

\

It has been observed by various workers that even low dimensional systems
such as the two-qubit system still give surprises. We have found this to be
the case in our investigation of factorizability of bases. Preliminary
investigations have suggested that in more complicated systems, such as a
bi-partite system of the form $\mathcal{H}\equiv \mathcal{H}%
_{1}^{(3)}\otimes \mathcal{H}_{2}^{(2)}$, the investigation of
factorizability of bases becomes harder rapidly because of an increased
number of combinatorial possibilities. It is not clear at this stage for
example whether the non-existence of any type $\left( 1,3\right) $ basis or
operator in the two qubit case has analogues in higher dimensional systems.
We have not to date found any type $(1,5)$ basis for $\mathcal{H}\equiv
\mathcal{H}_{1}^{(3)}\otimes \mathcal{H}_{2}^{(2)}$. A proof that one does
not exist has not been found yet, although it appears intuitively obvious.
Essentially, the single entangled basis element $\eta $ in such a basis type
would be of the form $\eta =\psi _{1}\otimes \phi _{1}+\psi _{2}\otimes \phi
_{2}+\ldots $, in obvious notation, such that $(\psi _{1},\psi _{2})_{1}=0$,
etc, but because $\eta $ has to be orthogonal to the subspace spanned by the
five mutually orthogonal and separable basis vectors $\eta _{1},\eta
_{2},\ldots ,\eta _{5}$, there is simply no ``space'' for such an $\eta $ to
exist. From another point of view, it can be seen that although any single
vector in a Hilbert space defines a one-dimensional subspace, even when it
is entangled, the property of entanglement itself requires at least two
dimensions for its operational definition. This argument leads us to
conjecture that in general, there is no type $(1,d_{1}d_{2}-1)$ basis for $%
\mathcal{H}\equiv \mathcal{H}_{1}^{(d_{1})}\otimes \mathcal{H}_{2}^{(d_{2})}.
$

\ $\;$

\subparagraph{\textbf{Acknowledgments:}}

J.E. acknowledges an EPSRC research studentship. We are grateful to A.
Sudbery for comments.

\end{document}